\documentclass[aps,prl,twocolumn,superscriptaddress, nofootinbib,  longbibliography, floatfix]{revtex4-1}
\usepackage{lineno}
\usepackage{graphicx} % needed for figs
\usepackage{dcolumn}  % needed for some tables
\usepackage{bm}    % for math
\usepackage{amssymb}  % for math
\usepackage{amsfonts, amsmath, amssymb, mathtools}
\usepackage{amsmath}

\usepackage{lipsum}
\usepackage[normalem]{ulem}
\usepackage{commath}
\usepackage{xcolor}  
\usepackage{placeins}
\newcolumntype{P}[1]{>{\centering\arraybackslash}p{#1}}
\usepackage[binary-units=true]{siunitx}
\usepackage{cleveref}
\RequirePackage{textcase}
\DeclareUnicodeCharacter{2009}{\,} 
\renewcommand{\selectlanguage}[1]{}

\begin{document}
% \linenumbers

\newcommand{\todo}[1]{{\color[rgb]{1.0,0.0,0.0}#1}} % gaps to be filled in red

\newcommand{\red}[1]{\textcolor{red}{#1}}
\newcommand{\todiscuss}[1]{\textcolor{teal}{#1}}
\newcommand{\tk}[1]{\textcolor{blue}{#1}}
\newcommand{\yg}[1]{\textcolor{purple}{#1}}

\raggedbottom

\renewcommand{\thesubsection}{\thesection.\arabic{subsection}}
\renewcommand{\thesubsubsection}{\thesubsection.\arabic{subsubsection}}

\title{Experimental demonstration of enhanced quantum tomography \\via quantum reservoir processing}

\author{Tanjung Krisnanda}
\email[Corresponding author: ]{tanjung@nus.edu.sg}
\author{Pengtao Song}
\author{Adrian Copetudo}
\author{Clara Yun Fontaine}
\affiliation{Centre for Quantum Technologies, National University of Singapore, Singapore 117543, Singapore}
\author{Tomasz Paterek}
\affiliation{School of Mathematics and Physics, Xiamen University Malaysia, Sepang 43900, Malaysia}
\affiliation{Institute of Theoretical Physics and Astrophysics, Faculty of Mathematics, Physics and Informatics, University of Gda\'{n}sk, Gda\'{n}sk 80-308, Poland}
\author{Timothy C. H. Liew}
\affiliation{School of Physical and Mathematical
Sciences, Nanyang Technological University, Singapore 637371, Singapore}
\author{Yvonne Y. Gao}
\email[Corresponding author: ]{yvonne.gao@nus.edu.sg}
\affiliation{Centre for Quantum Technologies, National University of Singapore, Singapore 117543, Singapore}
\affiliation{Department of Physics,
National University of Singapore, Singapore 117542, Singapore}
\date{\today}

\begin{abstract}
Quantum machine learning is a rapidly advancing discipline that leverages the features of quantum mechanics to enhance the performance of computational tasks. Quantum reservoir processing, which allows efficient optimization of a single output layer without precise control over the quantum system, stands out as one of the most versatile and practical quantum machine learning techniques. Here we experimentally demonstrate a quantum reservoir processing approach for continuous-variable state reconstruction on a bosonic circuit quantum electrodynamics platform. The scheme learns the true dynamical process through a minimum set of measurement outcomes of a known set of initial states. We show that the map learnt this way achieves high reconstruction fidelity for several test states, offering significantly enhanced performance over using a map calculated based on an idealised model of the system. This is due to a key feature of reservoir processing which accurately accounts for physical non-idealities such as decoherence, spurious dynamics, and systematic errors. Our results present a valuable tool for robust bosonic state and process reconstruction, concretely demonstrating the power of quantum reservoir processing in enhancing real-world applications.
\end{abstract}

\maketitle

%%%%%%%%%%%%%%%%%%%%%%%%%%%%%%%%%%%%%%%%%%%%%%%%%%%
\section*{Introduction}
%%%%%%%%%%%%%%%%%%%%%%%%%%%%%%%%%%%%%%%%%%%%%%%%%%%

In the classical domain, reservoir computing is a machine learning approach that leverages a network of randomly coupled nodes, known as the reservoir~\cite{lukovsevivcius2012practical}. 
This method does not require control over the reservoir, with optimization efforts focused solely on a single output layer.
The elimination of stringent hardware requirements makes reservoir computing an effective strategy for a wide range of tasks and particularly attractive for practical implementation.
For instance, reservoir computing has been successfully applied to time series prediction~\cite{brunner2013parallel}, speech recognition~\cite{vandoorne2014experimental,larger2017high}, and the characterization~\cite{jaeger2004harnessing,pathak2018model} and prediction~\cite{rafayelyan2020large} of chaos. 
Its hardware-friendly nature has enabled its implementation on various platforms, including photonic systems~\cite{vandoorne2014experimental,larger2017high,brunner2013parallel}, soft machines~\cite{nakajima2015information}, memristor arrays~\cite{kudithipudi2016design,du2017reservoir,midya2019artificial}, origami~\cite{bhovad2021physical}, magnetic systems~\cite{dale2024reservoir,allwood2023perspective}, and exciton-polariton lattices~\cite{ballarini2020polaritonic,mirek2021neuromorphic}.

Thanks to its simplicity and effectiveness, reservoir computing strategies have also been extended into the quantum domain~\cite{fujii2017harnessing,ghosh2019quantum,mujal2021opportunities,fujii2021quantum,martinez2021dynamical,innocenti2023potential,vetrano2025state}, and can be found under different names: quantum reservoir computing, quantum extreme learning machines, or quantum reservoir processing (QRP).
% which we will refer to as quantum reservoir processing (QRP). 
Here, schemes typically aim to use variational machine learning techniques in quantum systems that do not require precise control or optimization, making QRP an attractive candidate for solving complex tasks on current noisy intermediate-scale quantum devices. Proposed applications include both classical tasks, such as image recognition~\cite{xu2021superpolynomial} and time series prediction~\cite{kutvonen2020optimizing,mujal2023time}, as well as quantum-specific tasks of quantum state reconstruction~\cite{ghosh2020reconstructing,angelatos2021reservoir}, metrology~\cite{krisnanda2022phase}, state preparation~\cite{ghosh2019quantumstateprep, suprano2021dynamical}, circuit compression~\cite{ghosh2021realising}, and universal reservoir computing~\cite{nokkala2021gaussian}. Until very recently, experimental demonstrations have been limited to only classical tasks~\cite{negoro2018machine,dudas2023quantum}. In Refs.~\cite{suprano2024experimental,zia2025quantum}, experimental works have shown property estimation of up to two polarization encoded qubits.

One powerful quantum use case of QRP is the seemingly simple yet demanding task of a quantum state and process reconstruction, which are essential tools in quantum information technology. Accurate reconstructions provide complete knowledge on both the creation of complex resource states in the device as well as the dynamical processes they undergo. This task is particularly difficult for continuous-variable (CV) systems, which are prominent candidates for many quantum informational processing tasks due to their ability to host robust quantum resource states~\cite{larsen2019deterministic, asavanant2019generation, cochrane1999macroscopically, gottesman2001encoding, michael2016new} and realize hardware-efficient error correction~\cite{campagne2020quantum, ofek2016extending, ni2023beating, sivak2023real}. Thus, an accurate and efficient strategy to extract information from these complex quantum systems is highly desirable and remains an area of active investigation~\cite{landon2018quantitative, he2024efficient, chakram2022multimode, shen2016optimized, krisnanda2024demonstrating}. 
Machine learning techniques have already been applied in experimental quantum tomography~\cite{palmieri2020experimental,tiunov2020experimental}, however these methods are based on classical neural networks.
For example, the technique in Ref.~\cite{palmieri2020experimental} uses the network composed of almost $700$ nodes and $7000$ states for training to accurately reconstruct states in six-dimensional space.
In contrast, our implementation leverages a quantum system where the same task is achieved with only a single qubit-cavity system as the resource and a minimalistic training set of $36$ elements, highlighting both the efficiency and potential scalability of this strategy.

\begin{figure}[tbh!]
\centering
\includegraphics[width=0.5\textwidth]{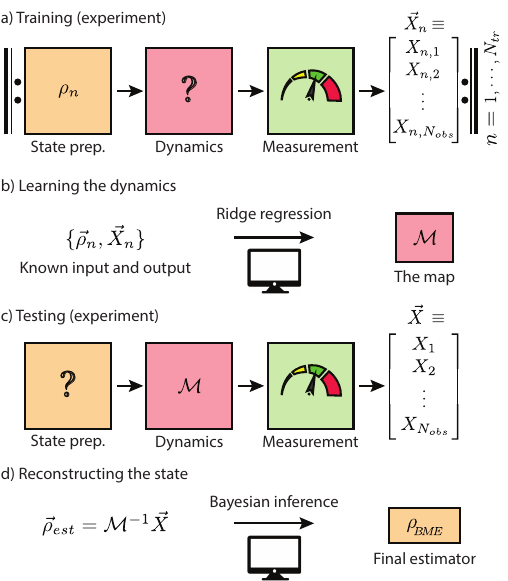}
\caption{{\bf Quantum reservoir protocol for state and process tomography.}
The basic idea of quantum reservoir processing is to perform simple measurements on a quantum system with unknown dynamics and post-process the results to achieve a particular task. (a) First, we reconstruct the system's dynamics (process tomography) in the training phase. This is done by preparing $N_{tr}$ known quantum states, subjecting them to an unknown dynamical process (described by the map to be learnt), followed by the measurement of $N_{obs}$ observables.
(b) The relation between the input state $\vec \rho_n$ (vectorised) and its corresponding measurement outcomes $\vec X_n$ allows for determination of the dynamical map $\mathcal{M}$.
Having reconstructed the dynamics, we move on to the problem of state tomography.
(c) We prepare a previously unseen state that undergoes the same dynamics and measurements.
(d) The learnt map is then inverted to estimate the input state $\vec \rho_{est}$ given the measured observables $\vec X$. Bayesian inference is subsequently applied to produce the most accurate physical estimate $\rho_{BME}$ of the density matrix. }
\label{fig: Fig1}
\end{figure}
 
Our QRP-based tomography scheme is implemented on an archetypal bosonic circuit quantum electrodynamics (cQED) platform to demonstrate robust state and process reconstructions of a CV system. Following the QRP methodology, we do not require precise control over the system's dynamics and reconstruct it by processing only the measurement outcomes of a known set of input states.
This then allows us to reconstruct unknown input quantum states, which we experimentally implement 
up to a truncation dimension of $D=6$.  
The data demonstrates that learning effectively accounts for imperfections such as dynamical errors, decoherence, and systematic measurement errors inherent in the systems. The learnt map achieves enhanced state reconstruction fidelities $>91\%$, in contrast to the $59\%$ obtained through the standard map computed based on an idealised model of the system. This QRP-based technique provides a simple yet powerful strategy for the accurate and efficient extraction of quantum information in CV systems. Our results also serve as a compelling demonstration of the power and efficacy of QRP as a general tool for quantum information processing on real hardware, subjected to both known physical imperfections and spurious, unknown dynamics.

\section*{The general protocol}
We begin by explaining in detail the steps in the protocol described in Fig.~\ref{fig: Fig1}. As the Hilbert space dimension of a bosonic mode is theoretically infinite, practical constraints necessitate setting a truncation dimension $D$, within which an \emph{arbitrary} bosonic state can be feasibly reconstructed.  
Each $D$-dimensional input state, in general mixed,
is defined by its $D^2-1$ independent real parameters. This sets the theoretical minimum number of single-outcome measurements (observables) required for state reconstruction at $N_{obs}=D^2-1$, a standard we adhere to in this study.

The dynamics is embodied by a completely positive and trace preserving map, which accommodates a wide range of scenarios, including those describable by the Lindblad master equation. Consequently, the input state $\rho_n$, represented by its $D^2-1$ parameters in a vector form $\vec Y_n$, is directly related to the observables $\vec X_n$ through a linear equation $\vec X_n=M\vec Y_n +\vec V$, where $M$ and $\vec V$, respectively, denote a matrix and constant vector~\cite{krisnanda2023tomographic}. 
For simplicity, this can also be written as $\vec X_n=\beta [1;\vec Y_n]$, where $[1;\vec Y_n]$ is a vector with the first entry equal to $1$, and $\beta \equiv [\vec V,\:\:M]$ is a $(D^2-1)\times D^2$ matrix representing the ``process" map. Note that $\beta$ includes both the dynamics and the measurement, see Appendix~\ref{apx-process-tomo}. To reconstruct the map, we perform learning with ridge regression (detailed in the Appendix Eq.~(\ref{EQ_ridgeregression})), requiring at least $D^2$ different input states ($N_{tr}=D^2$), a criterion consistently applied throughout our experiments.

During the state reconstruction phase, an arbitrary input state is generated and subjected to the same dynamical processes and measurements as in the learning phase. These measurement results are processed through an output layer, which effectively acts as the inverse of the map~\cite{krisnanda2023tomographic}, formulated as $\vec Y_{est}=M^{-1}(\vec X-\vec V)$, where $M^{-1}$ denotes the matrix inverse. However, due to imperfections in experiments, $\vec Y_{est}$ often fails to represent a physical state. To address this issue, we employ Bayesian inference as an additional step to obtain the most accurate physical estimator $\rho_{BME}$ based on the data. A more detailed discussion is presented in the Appendix.

\section*{Implementation in cQED}

\begin{figure}[b]
\centering
\includegraphics[width=0.5\textwidth]{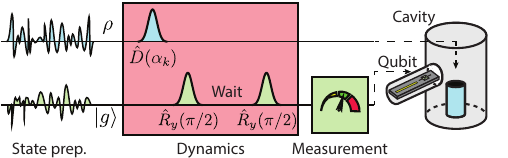}
\caption{{\bf Implementation of QRP in cQED.} A conventional bosonic cQED setup includes a 3D cavity that hosts the bosonic mode (reservoir), complemented by an ancillary qubit that assists in controlling and measuring the cavity state. An arbitrary input state $\rho$ can be prepared using the GRAPE method~\cite{heeres2017implementing}. Our target dynamical process is a sequence of a displacement operation $\hat D(\alpha_k)$ followed by parity mapping of the cavity state to a qubit state, which is subsequently measured. The parity mapping utilizes a standard $\pi/2$-wait-$\pi/2$ Ramsey sequence. In practice, this dynamics is not realised perfectly and the machine learning method is used to reconstruct the actual process.} 
\label{fig: Fig2}
\end{figure}

We implement the protocol depicted in Fig.~\ref{fig: Fig1} on a standard bosonic cQED platform. The device, as shown in Fig.~\ref{fig: Fig2}, features a long-lived 3D cavity that hosts the bosonic mode. It acts as the reservoir that is dispersively coupled to an ancillary transmon qubit, which introduces the necessary nonlinearity for state preparation and measurement of the cavity state. The Hamiltonian for the qubit-cavity system in a rotating frame with the frequencies of the qubit and cavity drives is expressed as 
\begin{eqnarray}
\hat H/\hbar &=& -\chi \hat c^{\dagger}\hat c |e\rangle \langle e| +\epsilon_{c}(t)\hat c + \epsilon_{c}^*(t) \hat c^{\dagger} \nonumber \\
&&+\epsilon_{q}(t) \hat \sigma_- + \epsilon_{q}^*(t) \hat \sigma_+,\label{EQ_Hamiltonian}
\end{eqnarray}
where $\hat c$ ($\hat c^{\dagger}$) represents the annihilation (creation) operator for the bosonic mode, $\hat \sigma_- = | g \rangle \langle e|$ ($\hat \sigma_+ = | e \rangle \langle g|$) the annihilation (creation) operator for the qubit with $|g\rangle$ ($|e\rangle$) as its ground (excited) state, and $\chi/2\pi\approx 1.42$~MHz the qubit-cavity dispersive coupling strength. The $\epsilon_j(t)$ denotes the time-dependent controlled drives with the subscript $j$ indicating the qubit ($q$) or cavity ($c$). A more detailed Hamiltonian including the higher order (weaker) terms is presented in the Appendix. In addition, the main decoherence mechanisms relevant for this experiment are qubit energy decay, qubit dephasing, and cavity energy decay. These properties are characterised via standard cQED experiments and detailed in the Appendix.

For the training phase, we generate $D^2$ different pure input states $\{|\psi_n\rangle\}$: $D$ Fock states ($|0\rangle, |1\rangle,\cdots,|D-1\rangle$) and $D^2-D$ of their superpositions $(|l\rangle+e^{i\Phi}|m\rangle)/\sqrt{2}$, where $l<m=0,1,\cdots,D-1$ and $\Phi=\{0,\pi/2\}$. This is done by optimizing $2~\mu$s-long time-dependent drives $\epsilon_j(t)$ in Eq.~(\ref{EQ_Hamiltonian}) using the Gradient Ascent Pulse Engineering (GRAPE) method~\cite{heeres2017implementing}. 

To execute the protocol, our target dynamics is composed of displacement operators $\{\hat D(\alpha_k)\equiv \exp(\alpha_k \hat c^{\dagger}-\alpha_k^* \hat c)\}_{k=1}^{D^2-1}$ applied to the cavity state, each followed by the same parity mapping to the qubit state (Fig.~\ref{fig: Fig2}). This mapping is conducted using the standard Ramsey technique, consisting of a $\pi/2$-wait-$\pi/2$ sequence~\cite{lutterbach1997method, bertet2002direct}, where the $\pi/2$-pulse is a $\pi/2$ qubit rotation about the $\hat y$ axis on the Bloch sphere, and during the wait time $t_w=\pi/\chi$, the dispersive coupling invokes a conditional phase gate $\hat C=|g\rangle \langle g| \otimes \mathbb{I} +|e\rangle \langle e| \otimes e^{i\pi \hat c^{\dagger}\hat c}$. Ideally, this sequence converts the parity $P$ of the cavity state before the sequence to the qubit state through $P=-\langle \hat \sigma_z \rangle$, where $\hat \sigma_z$ is the qubit Pauli-$z$ operator. The parity then constitutes the set of observables $\vec X_{n}$ in our protocol. In practice, the mean value $\langle \hat \sigma_z \rangle$ is obtained via single-shot qubit measurements averaged over 1000 repetitions, without any correction that may come from systematic measurement errors. Notably, the choice of displacement amplitudes $\{\alpha_k \}_{k=1}^{D^2-1}$ for a given dimension $D$ is not random. These amplitudes are optimized using a gradient descent method to yield robust dynamics for distinguishing arbitrary states with the minimum number of measurements, see details in the Appendix.

\section*{Results}
\subsection*{Quantum process reconstruction}

While the experimental implementation of the displacements on the cavity is robust, the subsequent parity mapping is known for its inherent errors that cannot be completely corrected. As such, native state reconstruction procedures with the computed ``idealised map", denoted by $\beta_I$ (detailed in the Appendix Eq.~(\ref{EQ_linearequation})), will yield erroneous outcomes~\cite{krisnanda2024demonstrating}. In contrast, the QRP strategy thrives in capturing processes that involve complex or unknown dynamics by design. 
We call the map obtained by 
implementing the protocol depicted in Fig.~\ref{fig: Fig1}  
the ``learnt map" and denote it by $\beta_L$. 
Its most important feature is the elimination of the dynamical error from the parity mapping as will be explained below.

\begin{figure}[bt]
\centering
\includegraphics[width=0.5\textwidth]{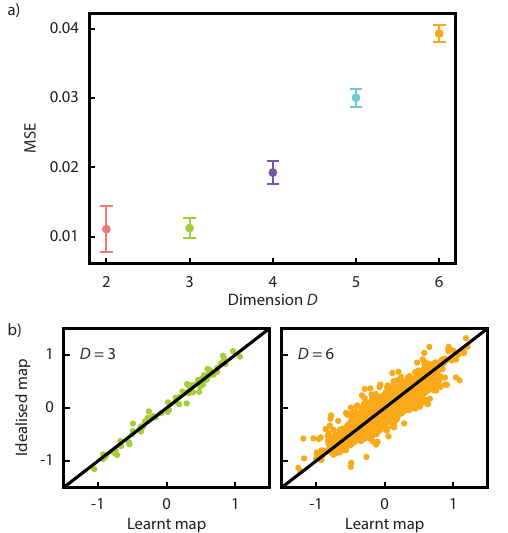}
\caption{{\bf Deviation between the idealised and learnt maps.} (a) The element-wise mean square error (MSE), averaged over all elements of $\beta_{I}$ and $\beta_{L}$, for various truncation dimensions $D$.
Error bars are from bootstrapping. Deviation between the maps grows with the dimension. (b) Elements of $\beta_I$ plotted against those of $\beta_L$ for exemplary truncation dimensions $D=3$ and $6$. Diagonal lines indicate points of equality.} 
\label{fig: Fig3}
\end{figure}

To benchmark the performance of QRP in obtaining an accurate map, we decouple the imperfections that stem from state preparation by first independently simulating the initial state with real hardware parameters. In other words, we construct the set $\{\vec Y_n(\rho_n),\vec X_n$\}, with $\rho_n$ being the density matrix simulated from GRAPE pulses taking into account decoherence affecting the system.
We note that the simulated density matrices $\{\rho_n\}$ are close to the ideal states, $\{|\psi_n\rangle \}$, with average fidelity $\approx 0.97$ which is limited by the finite transmon decoherence time ($\approx15$\,$\mu$s) during the GRAPE pulse ($\approx2$\,$\mu$s). The $D^2-1$ real parameters that form $\vec Y_n$ are then taken from the density matrix $\rho_n$. We take the first $D-1$ diagonal elements, since the normalisation condition infers the last element, as well as the $D^2-D$ real and imaginary off-diagonal (upper triangular) elements.

We present a comparison of the idealised and learnt maps across various truncation dimensions $D$ in Fig.~\ref{fig: Fig3}. Our findings indicate that the idealised map significantly deviates from the learnt map, revealing the breakdown of the perfect parity assumption. Notably, the deviation, measured as the element-wise mean square error (MSE) $\sum_{ij}(\beta_I-\beta_L)_{ij}^2/(D^2(D^2-1))$ averaged over all $D^4-D^2$ elements, increases with truncation dimension (Fig.~\ref{fig: Fig3}a). The primary source of error contributing to this discrepancy stems from coherent dynamical errors during the parity mapping process. Specifically, the $\pi/2$ pulses are not perfect, as the dispersive qubit-cavity coupling, which is parasitic during these pulses, remains active. The higher the dimension of the states we want to reconstruct, the higher the coherent error~\cite{krisnanda2024demonstrating}. This imperfection, if not learnt, significantly impacts the accuracy of state reconstruction using the idealised map as we will see below.

\begin{figure}[b!]
\centering
\includegraphics[width=0.5\textwidth]{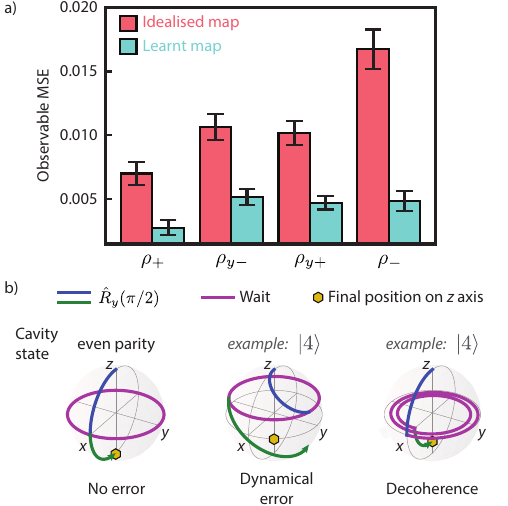}
\caption{{\bf Errors of the observables for four kitten states.} 
(a) The observable MSE under the assumption of idealised map (pink bars) and learnt map (cyan bars). The errors are averaged over all 
$D^2-1$ observables. 
It is clear that the learnt map has superior performance. The observed error reduction highlights the effectiveness of the learning process in addressing the majority of errors, including dynamical errors, decoherence, and systematic errors. Error bars are obtained from bootstrapping. 
(b) Illustration of the errors from the parity mapping sequence on the qubit's Bloch sphere. The left-most figure shows idealised dynamics of any even parity cavity state. The always-on dispersive coupling constitutes dynamical errors during both $\pi/2$ pulses (blue and green curves). The decoherence, dominantly qubit dephasing, results in inward-spiraling trajectory (purple line) mainly during the wait time.} 
\label{fig: Fig4}
\end{figure}

\begin{figure*}
\centering
\includegraphics[width=1.0\textwidth]{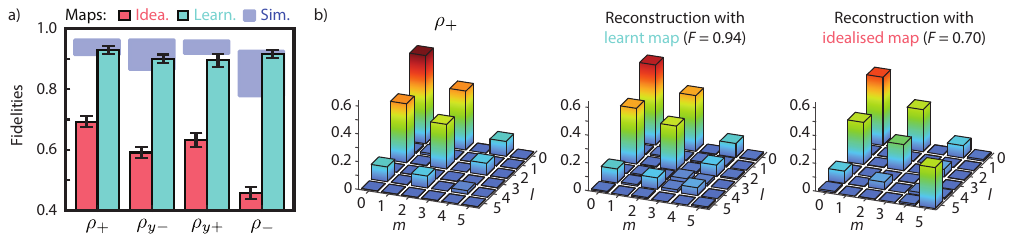}
\caption{{\bf Reconstruction of kitten states.} 
(a) The fidelities of reconstructed kitten states within $D=6$ using three different methods: the idealised map (pink bars), the learnt map (cyan bars), and the simulated map (blue-shaded area). Error bars are derived from bootstrap analysis. The simulated map is generated by simulating the actual experimental pulses using precise dynamical parameters. The highest points within the blue area indicate the best dynamical parameters, while the lowest points reflect a $2\%$ error in the dispersive coupling $\chi$ and $50\%$ error in the self-Kerr and second order dispersive coupling.
(b) A tomographic plot of the reconstructed density matrix of the even kitten state $\rho_+$ in the Fock basis.} 
\label{fig: Fig5}
\end{figure*}

We evaluate the performance of both the idealised and learnt maps by testing them on states distinct from those used in the training phase. Inspired by cat states, we generated four kitten states of the form $|\psi _{\pm} \rangle \equiv |\alpha \rangle \pm |-\alpha \rangle$ and $|\psi _{y\pm} \rangle \equiv|\alpha \rangle \pm i |-\alpha \rangle$, with $\alpha = 1$, normalized appropriately.
The corresponding states simulated from GRAPE pulses are denoted by $\rho_{\pm}$ and $\rho_{y\pm}$, respectively.
As these states are well contained within $D=6$, we use the $D=6$ maps. The states were prepared using the numerically optimized pulses of a $2~\mu$s duration using the GRAPE method. The same dynamics were then applied to each state, followed by measurements of their respective observables $\vec X$. This testing allows us to assess the robustness and adaptability of the maps to new quantum states, providing a thorough evaluation of their practical applicability in diverse scenarios.

We analyze the discrepancy between the measured and the expected observables, where the latter is computed based on the idealised or learnt map.
Specifically, we define observable MSE as:
\begin{equation}
 \frac{1}{D^2-1}\sum_j (\vec{X} - \beta \left[ \begin{array}{c}
         1 \\
         \vec{Y} 
    \end{array}\right])_j^2,
\end{equation}
where \(\beta\) represents the map being tested and the mean is over all the $D^2-1$ individual observables. See Appendix~\ref{apx_observable_error} for analysis of error for different observables. The observable MSE is illustrated in Fig.~\ref{fig: Fig4}a for the four kitten states. Our results clearly demonstrate that the error associated with the idealised map is consistently higher than that of the learnt map. 

Experimental errors manifest in several forms, such as dynamical errors from inaccurate operations, decoherence due to information loss to the environment, and both systematic and random measurement errors. In our case, the first two constitute significant errors during the parity mapping process, as illustrated in Fig.~\ref{fig: Fig4}b. Notably, the learning protocol addresses all these imperfections except for the random measurement errors, in contrast to the idealised map that neglects all error types. As a result, the impact of errors on the observables is suppressed when using the learnt map.

\subsection*{Quantum state reconstruction}

We directly apply the state reconstruction protocol outlined in Fig.~\ref{fig: Fig1}d to the measured results \(\vec{X}\). To evaluate the quality of the reconstructed state, we calculate its fidelity to a target state using the formula:
\begin{equation}
F = \left(\text{tr} \sqrt{\sqrt{\rho_{\text{tar}}} \rho_{\text{BME}} \sqrt{\rho_{\text{tar}}}}\right)^2,
\end{equation}
where \(\rho_{\text{BME}}\) represents the estimated state obtained from Bayesian inference, and \(\rho_{\text{tar}}\) is the target state. 

The fidelities of the four kitten states processed with the learnt map significantly exceed those computed using the idealised map, as shown in Fig.~\ref{fig: Fig5}a. For further comparison, we also constructed a map that incorporates decoherence effects and realistic dynamical pulses, which necessitates precise knowledge of the Hamiltonian parameters, including higher order terms, and the dynamical pulses (see the Appendix). 
This map is called ``simulated map'' in Fig.~\ref{fig: Fig5}a.
To account for limited measurement resolutions, which are in the order of $\sim$~kHz in real experiments, we introduced into the simulated map a \(2\%\) error in the dispersive coupling \(\chi\) and \(50\%\) in the higher order terms (both the self-Kerr and second order dispersive coupling) to demonstrate the effects of these potential inaccuracies on the reconstruction fidelity. The blue shaded region in Fig.~\ref{fig: Fig5}a illustrates these fidelities. The highest points correspond to the cases where inaccuracies in the system parameters reach a minimum, while the lowest points reflect the scenarios where the parameters used in the simulation differ most significantly from the properties of the actual device. We present the reconstructed density matrices of the even kitten state in Fig.~\ref{fig: Fig5}b to illustrate the performance of the learnt and idealised maps.

We note that, in general, the learnt map significantly outperforms the idealised map due to the capability of QRP to accurately account for complex dynamics in the system. While the simulated map offers the highest fidelities with certain combinations of system parameters, its performance is highly volatile and deteriorates rapidly even with a small range of parameter mismatches at the kHz level. Going beyond this level of precision in practice is technically challenging and resource intensive. In contrast, the learnt map achieves reconstruction fidelities comparable to the best case scenarios of the simulated maps and affords much better stability.

\section*{Conclusion}

We have successfully implemented a QRP architecture on a bosonic cQED device and demonstrated effective bosonic state and process reconstruction with significantly enhanced performance compared to standard methods. We obtained an accurate process map that circumvents the flaws of the typical approach of an idealised map computed naively based on an assumed model of the system with perfect operations. Furthermore, we tested both the idealised and the learnt maps on a set of kitten states, and showed that state reconstruction fidelities are significantly enhanced by using the learnt map obtained through QRP. Our study provides a simple yet powerful strategy to implement robust quantum tomography in CV system. More generally, this result also highlights the compelling advantages of QRP-based techniques in realizing practical machine learning applications on current quantum hardware.

\section*{Data and code availability}
All the data and codes needed to evaluate the conclusions of the paper are available on GitHub: \url{https://github.com/tkrisnanda/Tomography_QRP}.

\section*{Acknowledgments}
We acknowledge funding support from the Ministry of Education (MOE-T2EP50121-0020) and the National Research Foundation (NRFF12-2020-0063), Singapore. A.C. and C.Y.F. acknowledge the funding support from the CQT Ph.D scholarship.

\section*{Competing interests}
The authors declare no competing interests.

% \section*{Additional information}
% Supplementary information for this article is available.

\appendix
\section{Learning with ridge regression}
The learnt map, denoted as \(\beta_L \equiv [\vec V,\: M]\), is derived from the training dataset \(\{\vec Y_n, \vec X_n\}_{n=1}^{N_{tr}}\) using ridge regression. 
Both $\vec Y_n$ and $\vec X_n$ are $(D^2-1)\times 1$ vectors as we consider $D^2-1$ real parameters of the input state and $D^2-1$ measured observables, respectively. We also note that $\vec V$ is a $D^2-1$ vector and $M$ is a $(D^2-1)\times (D^2-1)$ square matrix. Consequently, $\beta_L$ is a $(D^2-1)\times D^2$ matrix consisting of the vector $\vec V$ as its first column and matrix $M$ filling the rest of its elements. Recall that a general CPTP map of dimension $D$ has $D^4-D^2$ real parameters, which is inline with the dimension of our map $\beta_L$.

We define the output matrix \(\mathcal{X}\) and the input matrix \(\mathcal{Y}\) as follows:
\begin{eqnarray}
    \mathcal{X} &\equiv& \left[\begin{array}{cccc}
        \vec X_1 & \vec X_2 & \cdots & \vec X_{N_{tr}}
    \end{array}\right] \nonumber \\
    \mathcal{Y} &\equiv& \left[\begin{array}{cccc}
        1 & 1& \cdots& 1 \\
        \vec Y_1 & \vec Y_2& \cdots& \vec Y_{N_{tr}}
    \end{array}\right].
\end{eqnarray}
The map \(\beta_L\) is computed using the equation:
\begin{equation}
    \beta_L = \mathcal{X}\mathcal{Y}^T(\mathcal{Y}\mathcal{Y}^T+\nu \mathbb{I})^{-1}\label{EQ_ridgeregression}
\end{equation}
where \(\nu\) is the regularization coefficient selected to optimize the balance between overfitting and underfitting from noisy data.

The case of $\nu=0$ represents linear regression. Here, we can see that the number of training states required is at least $D^2$ because to solve Eq.~(\ref{EQ_ridgeregression}) with $\nu=0$, one has to ensure the invertibility of the real matrix $\mathcal{Y}\mathcal{Y}^T$. This means that $N_{tr}$ has to be at least $D^2$, since $\text{det}(\mathcal{Y}\mathcal{Y}^T)=0$ for any rectangular matrix $\mathcal{Y}$ having fewer columns than rows.

\section{Optimized displacements $\{\alpha_k\}$ and idealised map $\beta_I$}
In this section, we describe the selection process for the set of displacement amplitudes \(\{\alpha_k\}\) given a truncation dimension \(D\) for state reconstruction. We start with the assumption of perfect parity mapping of the cavity state to the qubit state. Consequently, the expectation value of the \(k\)th observable is expressed as:
\begin{eqnarray}
    X_{n,k} &=& \text{tr}(\hat P\hat D(\alpha_k)\rho_n \hat D^{\dagger}(\alpha_k))\nonumber \\
    &=&\text{tr}(\hat D^{\dagger}(\alpha_k) \hat P\hat D(\alpha_k)\rho_n ),\label{EQ_observable}
\end{eqnarray}
where \(\rho_n = |\psi_n \rangle \langle \psi_n|\) and \(\hat P=e^{i\pi \hat c^{\dagger}\hat c}\) is the parity operator. 

From this formulation, as represented in all \(\{X_{n,k}\}_{k=1}^{D^2-1}\), it becomes clear that we can establish a linear matrix equation 
\begin{equation}
    \vec X_{n}=\mathcal{M} \vec \rho_n,\label{EQ_mapprocess}
\end{equation} 
where \(\vec \rho_n\) is the vectorized density matrix with dimension $D^2\times1$, and \(\mathcal{M}\) is a $(D^2-1)\times D^2$ matrix derived from the operations \(\{\hat D^{\dagger}(\alpha_k) \hat P\hat D(\alpha_k)\}_{k=1}^{D^2-1}\), each truncated to the dimension \(D\). It is also noted that \(\vec \rho_n\) of the state \(\rho_n\) linearly correlates to its \(D^2-1\) real parameters through the equation \(\vec \rho_n=K\vec Y_n + \vec C\), where $K$ is a $D^2\times(D^2-1)$ matrix and $\vec C$ is a $D^2\times 1$ vector. This leads us to the linear relationship:
\begin{equation}
    \vec X_n = M \vec Y_n + \vec V,\label{EQ_linearequation}
\end{equation}
where \(M=\mathcal{M}K\) is now a $(D^2-1)\times (D^2-1)$ matrix and \(\vec V=\mathcal{M}\vec C\) a $(D^2-1)\times 1$ vector. This formulation helps derive the idealised map \(\beta_I=[\vec V, M]\) as mentioned in the main text. 

Furthermore, inverting Eq.~(\ref{EQ_linearequation}) allows for the estimation of the cavity state based on measured observables. To enhance robustness, the set \(\{\alpha_k\}\) is chosen such that the condition number of \(M\) is minimized, which is crucial as the condition number magnifies the relative error in the observable, influencing the relative error in the estimated state parameters~\cite{horn2012matrix}. We employ a gradient-descent method to optimize the selection of \(\{\alpha_k\}_{k=1}^{D^2-1}\) for each truncation dimension \(D\) of the cavity state. Our codes are available on GitHub.

Note that \(M\) is a square matrix (as we use minimal number $D^2-1$ of measured observables). For a more general scenario, where the number of measurements, $N_{obs}$, is not equal to $D^2-1$, one can use a pseudoinverse to arrive at $\vec Y_{est}=M^+(\vec X-\vec V)$, where $M^+=(M^{\dagger}M)^{-1}M^{\dagger}$ is the left pseudoinverse.
Generally, the condition of information completeness for state tomography is fulfilled if the determinant $\text{det}(M^{\dagger}M)\ne0$, which is the invertibility of the square matrix $(M^{\dagger}M)$, a condition on the existence of the pseudoinverse~\cite{krisnanda2023tomographic}. From this, it follows that the number of observables has to be greater than or equal to the number of parameters for the density matrix, $N_{obs}\ge D^2-1$, since $\text{det}(M^{\dagger}M)=0$ for any rectangular matrix $M$ having fewer rows than columns. This is to be expected, as single-valued observables with cardinality less than $D^2-1$ are not enough to parametrize a density matrix of dimension $D$.

\section{The process tomography}
\label{apx-process-tomo}
Process tomography usually refers to the reconstruction of a linear map $\Lambda$, acting on an input density matrix $\Lambda(\rho)$. Working within a truncation dimension $D$ in a matrix representation, an equivalent expression can be written as $\mathcal{M}\vec \rho$, where $\mathcal{M}$ is a $D^2-1 \times D^2$ to-be-reconstructed linear map and $\vec \rho$ is the vectorized density matrix.

In the ideal case, assuming perfect operations, it is clear from Eq.~(\ref{EQ_mapprocess}) and its equivalent form Eq.~(\ref{EQ_linearequation}) that a set of measured results $\vec X$ is a representation of the state being the output of $\Lambda(\rho)$, as in the usual process tomography. 
When experimental imperfections are prevalent, which in our case is dominated by the dynamical error and qubit dephasing during the parity mapping, Eq.~(\ref{EQ_observable}) is no longer valid. However, as the imperfections are describable by CPTP map, the linear relation between the results and input state is preserved
\begin{equation}
    \vec X =\mathcal{M}^{\prime} \vec \rho,
\end{equation}
where $\mathcal{M}^{\prime}$ is the to-be-reconstructed linear map that incorporates the \emph{true dynamics} of the system and measurement.
As explained below Eq.~(\ref{EQ_linearequation}), one reconstructs the map $\mathcal{M}^{\prime}$ from the learnt map $\beta_L=[\vec V, M]$ using $\mathcal{M}^{\prime}=MK^+$, where $K^+$ here is the right Moore-Penrose
pseudoinverse of $K$.

We note that the process map $\mathcal{M}^{\prime}$ includes the measurement. Our method can be easily adapted to perform process tomography of unknown dynamics, excluding the measurement. Suppose we want to determine a map $\Lambda$ that evolves a quantum state as $\rho_t=\Lambda(\rho)$. We can write an equivalent matrix form $\vec \rho_t = \mathcal{M}\vec \rho$. After using our parametrization ($\vec \rho = K \vec Y+\vec C$) on both the initial and final state, we arrive at
\begin{eqnarray}
    \vec Y_t &=& K^+\mathcal{M}K\vec Y + K^+\mathcal{M}\vec C-K^+\vec C\nonumber \\
    &\equiv& \Phi \vec Y + \vec Q,
\end{eqnarray}
where we have defined a matrix $\Phi\equiv K^+\mathcal{M}K$, a vector $\vec Q = K^+\mathcal{M}\vec C-K^+\vec C$, and $K^+$ here is the left pseudoinverse. Obtaining $\Phi,\vec Q$ is equivalent to obtaining the map $\Lambda$.

In experiments, one would perform state tomography to get $\vec Y_t$, by measuring observables $\vec X_t$. For example, using our method, we have the relation $\vec X_t=M\vec Y_t+\vec V$, where now $M$ and $\vec V$ are known. Therefore
\begin{eqnarray}
    \vec X_t&=&M\vec Y_t+\vec V\nonumber \\
    &=&M (\Phi \vec Y + \vec Q)+\vec V\nonumber \\
    &\equiv&\Gamma\vec Y +\vec R,\nonumber
\end{eqnarray}
where we have defined a matrix $\Gamma\equiv M \Phi$ and a vector $\vec R\equiv M\vec Q+\vec V$. We then obtain $\Gamma,\vec R$ from known sets of input states and observables $\{\vec Y_n,\vec X_{t,n}\}$. After that, the matrix $\Phi$ and vector $\vec Q$ are computed as $M^+\Gamma$ and $M^+(\vec R-\vec V)$, where $M^+$ is the left pseudoinverse.

\section{Observable errors}
\label{apx_observable_error}

\begin{figure}[b!]
\centering
\includegraphics[width=0.5\textwidth]{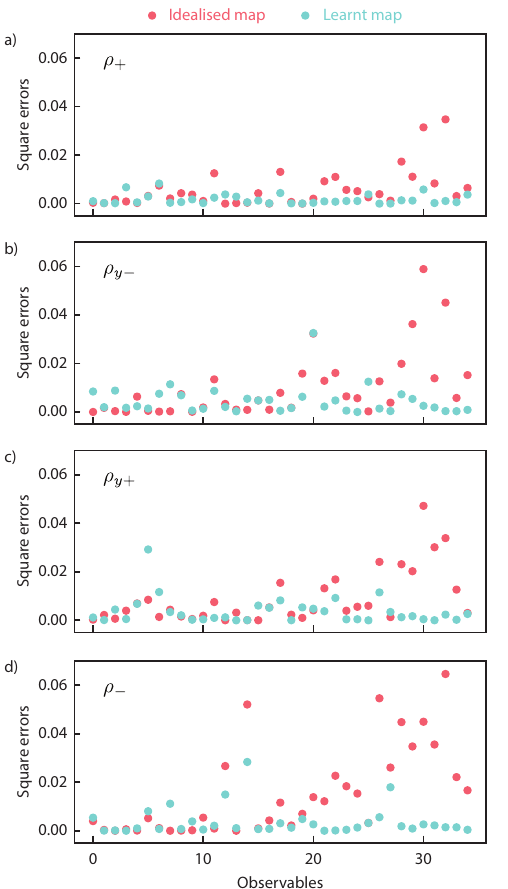}
\caption{{\bf Observable errors.} 
The square error for all $D^2-1=35$ observables for the cat states $\rho_{+}, \rho_{y-}, \rho_{y+},\rho_{-}$ respectively presented in panels (a), (b), (c), and (d). The observables have been arranged in the order of increasing displacement amplitude.} 
\label{fig: Fig6}
\end{figure}

In Fig.~\ref{fig: Fig6}, we present the square error
\begin{equation}
    (\vec{X} - \beta \left[ \begin{array}{c}
         1 \\
         \vec{Y} 
    \end{array}\right])_j^2,
\end{equation}
for all $j=1,2,\cdots,35$ observables of the cat states $\rho_{+}, \rho_{y-}, \rho_{y+},\rho_{-}$. The $x$-axis for the observables in Fig.~\ref{fig: Fig6} has been sorted in ascending order of the amplitude of the displacement $|\alpha|$ applied before the parity of the cavity is measured. The idealised map (pink circles)displays an increasing trend. This is because states displaced further away in phase space will experience more dynamical errors as already illustrated to some extend in Fig.~\ref{fig: Fig4}b. In contrast, the use of learnt map suppresses this error.

\section{The device}
\label{apx_sub:design-tools}
The experimental setup utilized in this study involves a standard bosonic cQED system, operating within the strong dispersive-coupling regime, as described in prior research~\cite{blais2004cavity, girvin2014circuit}. This system includes a superconducting microwave cavity linked to a transmon qubit, enhancing both control and readout capabilities, and is further connected to a planar readout resonator.

The storage cavity comprises a three-dimensional, high-Q coaxial \(\lambda/4\)-resonator with a cutoff frequency of approximately 600 MHz. Both the cavity and the coaxial waveguide that accommodates the qubit and the resonator are crafted from high-purity (5N) aluminum. To minimize manufacturing flaws, the outer layer of about 0.15 mm is chemically etched away.

The transmon qubit and the planar readout resonator are constructed by depositing aluminum on a sapphire substrate. The design is created using a Raith electron-beam lithography system on a HEMEX sapphire substrate, which is pre-cleaned using a 2:1 piranha solution for 20 minutes and subsequently coated with 800 nm of MMA and 250 nm of PMMA resist. The pattern is developed using a de-ionized water and isopropanol mixture at a 3:1 ratio. A PLASSYS double-angle evaporator is then employed to deposit two layers of aluminum, 20 nm and 30 nm thick, at angles of -25 and +25 degrees, respectively, with an intermediate oxidation step using an $85\%$ \(O_2\) and $15\%$ Argon mixture at 10 mBar for 10 minutes. Finally, the chip is diced using an Accretech machine and assembled into the waveguide using an aluminum clamp and indium wire to enhance thermalization.

\section{Hamiltonian parameters}

The bare Hamiltonian of our quantum system is given by
\begin{eqnarray}
\hat H_0/\hbar &=& \sum_{k=q,c,r} \omega_k \hat k^{\dagger} \hat k-\frac{\chi_{kk}}{2}\hat k^{\dagger} \hat k^{\dagger} \hat k \hat k \nonumber \\
&&-\chi_{cq} \hat c^{\dagger}\hat c \hat q^{\dagger}\hat q -\chi_{qr}  \hat q^{\dagger}\hat q \hat r^{\dagger}\hat r -\chi_{cr} \hat c^{\dagger}\hat c \hat r^{\dagger}\hat r \nonumber \\
&&-\chi_{cq}^{\prime} \hat c^{\dagger}\hat c^{\dagger}\hat c \hat c \hat q^{\dagger}\hat q,
\end{eqnarray}
where the indices \(k=q, c, r\) represent the transmon, cavity, and readout resonator, respectively. The terms \(\omega_k\) are the frequencies of each subsystem, \(\chi_{kk}\) and \(\chi_{kl}\) are the self-Kerr and cross-Kerr coefficients, respectively, and \(\chi_{cq}^{\prime}\) accounts for second-order interactions between the transmon and cavity. In this model, \(\hat k\) (\(\hat k^{\dagger}\)) signifies the bosonic annihilation (creation) operator for the respective system.
Note that considering the high self-Kerr of the transmon, it effectively functions as a two-level system (qubit) because higher energy levels are minimally occupied, thus allowing its dimension (and operators) to be truncated to \(D=2\). During the dynamics, the readout resonator remains in a vacuum state, which allows us to neglect its energy and Kerr coefficients. Also, the self-Kerr of the cavity and the second-order transmon-cavity interactions are weaker compared to the remaining terms. See Table~\ref{tab:Hparameters} for the list of parameters.

Further, controlled external drives can be applied as follows:
\begin{eqnarray}
\hat H_d/\hbar &=& \epsilon_{q}(t) \hat qe^{i\omega_{dq}t} + \epsilon_{q}^*(t) \hat q^{\dagger}e^{-i\omega_{dq}t} \nonumber \\
&&+\epsilon_{c}(t) \hat c e^{i\omega_{dc}t} + \epsilon_{c}^*(t) \hat c^{\dagger}e^{-i\omega_{dc}t} 
\end{eqnarray}
where \(\epsilon_j(t)\) represents the time-dependent strength of the transmon ($j=q$) and cavity ($j=c$) drives, and \(\omega_{dq}\) (\(\omega_{dc}\)) denotes the frequency of the drives applied to the transmon (cavity). When incorporated into the bare Hamiltonian and transformed into a rotating frame where the drive frequencies match the respective subsystem frequencies, it results in the effective Hamiltonian described in Eq.~(1) in the main text.

\begin{table}[h]
    \centering
    \begin{tabular}{|l|l|}
    \hline
    Parameters & Value \\
    \hline
     $\omega_q / 2\pi $ & 5.277 GHz \\
    \hline
     $\omega_c / 2\pi $ & 4.587 GHz \\
     \hline
     $\omega_r / 2\pi $ & 7.617 GHz \\
     \hline
     $\chi_{qq} / 2\pi $ & 175.3 MHz \\
     \hline
     $\chi_{cc} / 2\pi $ & 6 kHz \\
     \hline
     $\chi_{cq} / 2\pi $ & 1.423 MHz \\
     \hline
     $\chi_{qr} / 2\pi $ & 0.64 MHz \\
     \hline
     $\chi_{cr} / 2\pi $ & 2 kHz \\
     \hline
     $\chi_{cq}^{\prime} / 2\pi $ & 16 kHz \\
     \hline
     $T_{q1} $ & 85 $\mu$s \\
     \hline
     $T_{q\phi} $ & 15 $\mu$s \\
     \hline
     $T_{c1}  $ & 1 ms \\
     \hline
     $T_{r1}  $ & 2.1 $\mu$s \\
     \hline
    \end{tabular}
    \caption{List of Hamiltonian parameters.}    
    \label{tab:Hparameters}
\end{table}

\section{Operations and their duration}

In this section, we outline the operations and their duration as we execute the protocol illustrated in Fig.~2 of the main text. 
Initially, we conduct a measurement on the transmon qubit for preselection, proceeding only if the qubit is in its ground state ($\approx 97\%$), which is crucial for the subsequent state preparation phase. This preselection involves a single-shot measurement that takes 2 $\mu$s to populate and depopulate the readout resonator and an additional 1.6 $\mu$s of wait.

Following this, state preparation is carried out using the GRAPE method with optimized pulse amplitudes (\(\epsilon_{q}(t), \epsilon_{c}(t)\)) over a fixed duration of 2 $\mu$s. Displacements with amplitude \(\alpha_k\) are then executed using \(\epsilon_{c}(t)\) within 100 ns. Parity mapping involves a \(\pi/2\)-wait-\(\pi/2\) sequence, with each \(\pi/2\) pulse (using $\epsilon_{q}(t)$) lasting 64 ns and the wait period of 284 ns. A subsequent qubit measurement, also lasting 2 $\mu$s, is performed before allowing a 16 ms wait for the cavity state to decay to vacuum, prior to the next experimental run.

The expectation value of each observable is obtained with 1000 repetitions. For instance, when learning the map for \(D=6\), the total duration calculates to approximately 6 hours, primarily extended by the 16 ms waiting period for cavity state decay. 
A significantly faster process could be achieved by employing Q-switching, i.e., a microwave tone to swap the population from the cavity to the readout resonator, accelerating the decay of the cavity state~\cite{heeres2017implementing}, therefore, reducing the wait time to just tens of microseconds.
With this improvement, we project that the entire process of learning the map could be condensed to approximately 1~min.

\section{Simulated map}
We accurately emulate the retrieval of the learnt map using a classical computer, referring to this output as the simulated map. This simulation involves modeling the qubit-cavity system using the Lindblad master equation:
\begin{equation}
    \frac{d\rho}{dt}=-\frac{i}{\hbar}[\hat H, \rho]+\sum_m \hat J_m \rho \hat J_m^{\dagger}-\frac{1}{2}(\hat J_m^{\dagger}\hat J_m \rho +\rho \hat J_m^{\dagger}\hat J_m),
\end{equation}
where \(\hat H\) denotes the Hamiltonian consisting of the basic term \(\hat H_0\) and the driving terms \(\hat H_d\), which vary depending on the specific operation executed, and \(\hat J_m\) are the jump operators accounting for various decoherence mechanisms. Specifically, we include:
\begin{eqnarray}
    \hat J_1&=&\sqrt{1/T_{q1}}\:\: \hat q \nonumber \\
    \hat J_2&=&\sqrt{2/T_{q\phi}} \:\: \hat q^{\dagger}\hat q \nonumber \\
    \hat J_3&=&\sqrt{1/T_{c1}}\:\: \hat c,
\end{eqnarray}
representing qubit energy decay, qubit dephasing, and cavity energy decay, respectively.

This approach requires exact knowledge of Hamiltonian parameters and the precise application of experimental drive pulses. To explore the impact of inaccuracies, we conducted simulations incorporating a \(2\%\) error in the qubit-cavity dispersive coupling strength \(\chi\) and \(50\%\) error in the higher order terms (both the self-Kerr and second order dispersive coupling). The state reconstruction results obtained from the simulated map are displayed in Fig.~\ref{fig: Fig5}a.

\section{Bayesian inference}
The accurate determination of a quantum state from measurement outcomes is a critical task in quantum state reconstruction. Traditional methods like linear inversion (directly obtaining the estimated state from the estimated parameters $\vec Y_{est}$, which we refer to as $\rho_{LS}$) often result in non-physical states, as indicated by the appearance of negative eigenvalues. As a solution, Maximum Likelihood Estimation (MLE) has been commonly used, ensuring the estimated state is non-negative by maximizing the likelihood function that fits the observed data. However, MLE does not account for the uncertainty in state estimation and tends to result in zero eigenvalues (and therefore zero probabilities), which is an unrealistic assertion given finite amount of measured data~\cite{blume2010optimal,knips2015how}.

To address these limitations, our work implements Bayesian inference, a statistical approach that offers a more comprehensive view by incorporating a range of possible states and explicitly considering experimental uncertainties through Bayes' rule. Bayesian inference leverages prior knowledge about the system and updates this with experimental data to produce a posterior probability distribution. This distribution allows for estimating any function of \(\rho\) through numerical integration, which is computationally intensive. However, recent advances on Markov Chain Monte Carlo (MCMC) sampling algorithm have made this process very efficient. 

In particular, we implement the efficient Bayesian inference protocol proposed recently in Ref.~\cite{lukens2020practical} that utilizes pseudo-likelihood. 
The parameters we use when applying the protocol in Ref.~\cite{lukens2020practical} are as follows. We set \(\alpha=1\) to establish a uniform prior across all feasible physical density matrices, indicating no initial preference for any particular state. The variance of the pseudo-likelihood function centered on \(\rho_{\text{LS}}\) was defined as \(\sigma = 1/N\), where \(N = 1000 \cdot (D^2 - 1)\), to appropriately scale the influence of the data in updating our beliefs about the state. Additionally, we generated \(2^{10}\) MCMC samples, applying a thinning parameter of \(2^7\) to mitigate serial correlations within the sample chain. The protocol produces a Bayesian mean estimate that we use as the final estimator $\rho_{BME}$ for our state reconstruction.

\bibliography{Bibliography}    %use 

\end{document}